\documentclass[a4paper,preprint,superscriptaddress,nofootinbib,aps,pra,11pt]{revtex4-1}

\usepackage{graphicx} 
\usepackage{bm}
\usepackage{amssymb}
\usepackage{amsmath}
\usepackage{color}
\usepackage{cancel}
\usepackage{comment}
\usepackage{here}
\usepackage{float}
\usepackage{physics}
\usepackage{url}

\usepackage[toc,page]{appendix}

\begin{document}

\title{Revisiting the model for radiative neutrino masses with dark matter \\in the $\mathrm{U(1)}_{B-L}$ gauge theory}

\author{Shinya Kanemura}
\email{kanemu@het.phys.sci.osaka-u.ac.jp }
\affiliation{Department of Physics, Osaka University, Toyonaka, Osaka 560-0043, Japan}

\author{Yushi Mura}
\email{y\_mura@het.phys.sci.osaka-u.ac.jp }
\affiliation{Department of Physics, Osaka University, Toyonaka, Osaka 560-0043, Japan}

\author{Guohao Ying}
\email{ying@het.phys.sci.osaka-u.ac.jp }
\affiliation{Department of Physics, Osaka University, Toyonaka, Osaka 560-0043, Japan}

\preprint{OU-HET-1242}

\begin{abstract}
    The radiative seesaw model with gauged $\mathrm{U(1)}_{B-L}\times\mathbb{Z}_2$ extension is a well-motivated scenario which gives consistent predictions of active neutrino masses and the abundance of dark matter. 
    Majorana masses of right-handed neutrinos, the lightest of which can be identified as dark matter, are given by the spontaneous breaking of the $\mathrm{U(1)}_{B-L}$ gauge symmetry. 
    We revisit this model with the latest constraints from dark matter searches, neutrino oscillations, flavor experiments and collider experiments. 
    We explore the feasible parameter space of this model, and find that there are still allowed regions under the latest experimental constraints. 
    We present new viable benchmark scenarios for this model, i.e., the right-handed neutrino dark matter scenario and the scalar dark matter scenario. 
    We also mention the testability of these benchmark scenarios at future experiments.
\end{abstract}

\maketitle
\newpage

\section{Introduction}
Having the discovery of the Higgs boson at the Large Hadron Collider (LHC) in 2012, all the fundamental particles in the Standard Model (SM) of particle physics have been verified by experiments. 
However, there are some physics phenomena that the SM cannot explain. 
First, observations of the neutrino oscillation reveal that neutrinos have non-zero masses~\cite{Super-Kamiokande:1998kpq}. 
Second, the latest observation data provided by the Planck satellite~\cite{Planck:2018vyg} shows that approximately one quarter of the energy density in our universe is composed of non-baryonic and non-luminous matter, known as dark matter (DM). 
If the nature of dark matter is the weakly interacting massive particle (WIMP)~\cite{Lee:1977ua, Jungman:1995df}, the observed relic abundance can be satisfied with electroweak mass scale DM candidates. 
In addition, the universe exhibits matter and antimatter asymmetry.

A pioneer to explain the neutrino mass is the so-called seesaw mechanism~\cite{Minkowski:1977sc,Yanagida:1979as,Gell-Mann:1979vob,Ramond:1979py,Cheng:1980qt,Schechter:1980gr,Lazarides:1980nt,Mohapatra:1980yp,Magg:1980ut,Foot:1988aq}, which generates tiny neutrino masses at tree level.
For example, the type-I seesaw model~\cite{Minkowski:1977sc,Yanagida:1979as,Gell-Mann:1979vob,Ramond:1979py} can explain tiny neutrino masses with very heavy right-handed (RH) neutrinos or very small Yukawa couplings. 
On the other hand, in order to explain not only tiny neutrino masses, but also dark matter simultaneously, models with radiatively generated neutrino masses~\cite{Zee:1980ai,Zee:1985id,Babu:1988ki,Krauss:2002px,Tao:1996vb,Ma:2006km,Aoki:2008av} have often been studied, in which dark matter candidates are running in the loop of neutrino masses~\cite{Krauss:2002px,Tao:1996vb,Ma:2006km,Aoki:2008av}.    
A representative example is the model proposed by Tao and Ma~\cite{Tao:1996vb,Ma:2006km}, which can explain observed neutrino masses with TeV scale dark matter. 
However, this model does not explain the Majorana masses of RH neutrinos. 
The origin of Majorana masses of RH neutrinos can be explained in the radiative seesaw model with gauged $\mathrm{U(1)}_{B-L}\times\mathbb{Z}_2$ extension~\cite{Kanemura:2011vm}. 
RH neutrinos receive their masses through the spontaneous symmetry breaking (SSB) of the $\mathrm{U(1)_{B-L}}$ gauge symmetry~\cite{Okada:2010wd}. 
In Ref.~\cite{Kanemura:2011vm}, the observed DM relic abundance can be realized through the pair annihilation via the $s$-channel scalar exchange by the mixing between the SM Higgs field and the extra scalar singlet. 
The stability of DM, which is the lightest RH neutrino, is guaranteed by the unbroken $\mathbb{Z}_2$ symmetry. 
Tiny neutrino masses are generated radiatively via the one-loop induced operator. 
Some phenomenology of this model have been studied by Ref.~\cite{Borah:2018smz,Seto:2016pks}.

In this paper, we revisit the radiative seesaw model with gauged $\mathrm{U(1)}_{B-L}\times\mathbb{Z}_2$ extension~\cite{Kanemura:2011vm}, which was originally proposed before the discovery of the Higgs boson. 
Since experiments have made great progress in recent years, including flavor experiments~\cite{MEGII:2023ltw,SINDRUM:1987nra,BaBar:2009hkt,Belle:2021ysv,Hayasaka:2010np}, collider experiments~\cite{Electroweak:2003ram,ALEPH:2013htx,ATLAS:2019erb,ATLAS:2019fgd,CMS:2021ctt,CMS:2022eud,ATLAS:2022vkf,CMS:2022dwd}, neutrino oscillation measurements~\cite{ParticleDataGroup:2022pth} and dark matter searches~\cite{Planck:2018vyg,LZ:2022lsv}, it would be valuable to study whether there is still allowed parameter region where tiny neutrino masses and dark matter can be explained simultaneously under current constraints. 
We first discuss theoretical and experimental constraints on this model. 
We then consider neutrino masses, lepton flavor violation and DM physics in this model.
We show the relic density of DM at our benchmark point. 
The possible parameter space under current constraints is shown.
We then scan the allowed parameter space and give possible mass regions of DM.
Finally, we discuss the testability of this model at future experiments, including flavor experiments, collider experiments and DM detection experiments.

This paper is organized as follows. 
In section~\ref{sec:the_model}, we introduce the content of this model and give predictions of neutrino masses. 
In section~\ref{sec:constraint}, we show theoretical constraints and the latest experimental constraints for this model. 
In section~\ref{sec:CLFV}, we give predictions of lepton flavor violation in this model. 
In section~\ref{sec:DM}, we give the analysis of the lightest RH neutrino DM scenario and briefly discuss about the lightest inert scalar DM scenario. 
In section~\ref{sec:discussions}, we mention the testability of this model at future experiments. 
In section~\ref{sec:conclusion}, we summarize this work and give our conclusions.

\section{The model}\label{sec:the_model}
The radiative seesaw model with gauged $\mathrm{U(1)}_{B-L}\times\mathbb{Z}_2$ extension~\cite{Kanemura:2011vm} is a simple extension of the SM, in which we try to explain radiative generation of  neutrino masses with providing possible dark matter candidates. 
This model is an extension of the SM with three $\mathbb{Z}_2$ odd right-handed neutrinos $N_\alpha$  $(\alpha=1,2,3)$, a $\mathbb{Z}_2$ odd scalar  $\mathrm{SU(2)}_L$ doublet $\eta$, a scalar singlet $S$ and an electrically neutral $\mathrm{U}(1)_{B-L}$ gauge boson $Z^\prime$.
The Lagrangian is invariant under the gauge group $\mathrm{SU(3)}_C\times\mathrm{SU(2)}_L\times\mathrm{U(1)}_Y\times\mathrm{U(1)}_{B-L}$ with an unbroken $\mathbb{Z}_2$ discrete symmetry.
The particle content is shown in Table~\ref{tab:particle_content}.
\begin{table}[t]
    \centering
    \begin{tabular}{c|c c c c c c|c c c}
        \hline
        {}&$Q^i$&$u_R^i$&$d_R^i$&$L^i$&$e_R^i$&$\Phi$&$N_\alpha$&$\eta$&$S$\\
        \hline
        $\mathrm{SU(3)}_C$&$\mathbf{3}$&$\mathbf{3}$&$\mathbf{3}$&$\mathbf{1}$&$\mathbf{1}$&$\mathbf{1}$&$\mathbf{1}$&$\mathbf{1}$&$\mathbf{1}$\\
        \hline
        $\mathrm{SU(2)}_L$&$\mathbf{2}$&$\mathbf{1}$&$\mathbf{1}$&$\mathbf{2}$&$\mathbf{1}$&$\mathbf{2}$&$\mathbf{1}$&$\mathbf{2}$&$\mathbf{1}$\\
        \hline
        $\mathrm{U(1)}_Y$&$\frac{1}{6}$&$\frac{2}{3}$&$-\frac{1}{3}$&$-\frac{1}{2}$&$-1$&$\frac{1}{2}$&0&$\frac{1}{2}$&0\\
        \hline
        $\mathrm{U(1)}_{B-L}$&$\frac{1}{3}$&$\frac{1}{3}$&$\frac{1}{3}$&$-1$&$-1$&0&$-1$&0&$+2$\\
        \hline
        $\mathbb{Z}_2$&+&+&+&+&+&+&$-$&$-$&+\\
        \hline
    \end{tabular}
    \caption{Particle content and their quantum numbers.}
    \label{tab:particle_content}
\end{table}

The relevant interaction $\mathcal{L}_\mathrm{int}$ for our discussion is given by 
\begin{equation}
    \mathcal{L}_\mathrm{int} = \mathcal{L}^\mathrm{SM}_\mathrm{Yukawa}+\mathcal{L}_N-V(\Phi,\eta,S),
\end{equation}
where $\mathcal{L}^\mathrm{SM}_\mathrm{Yukawa}$ is the SM Yukawa interaction, and 
\begin{equation}
    \mathcal{L}_N=\sum_{\alpha=1}^3\qty(-\sum_{i=1}^3g_{i\alpha}\overline{L_i}\tilde{\eta}N_\alpha-\dfrac{y^R_\alpha}{2}\overline{N^c_\alpha}SN_\alpha+\mathrm{h.c.})
\end{equation}
with $\tilde{\eta}=i\tau^2\eta^*$, where $i,\alpha = 1,2,3$ are flavors of leptons.
Without loss of generality, the Yukawa coupling $y^R$ of RH neutrinos can be flavor diagonal. 
The Yukawa coupling among $\overline{L_i}\Phi N_\alpha$ is prohibited due to the unbroken $\mathbb{Z}_2$ symmetry. 
The active neutrinos remain massless at tree level, and obtain tiny masses at one-loop level.

The scalar potential $V(\Phi,\eta,S)$ in this model is given by
\begin{equation}
    \begin{aligned}
        V(\Phi,\eta,S) &= \mu_1^2\abs{\Phi}^2 + \mu_2^2\abs{\eta}^2 + \mu_S^2\abs{S}^2 + \dfrac{\lambda_1}{2}\abs{\Phi}^4 + \dfrac{\lambda_2}{2}\abs{\eta}^4 + \lambda_S\abs{S}^4\\
        &+ \lambda_3\abs{\Phi}^2\abs{\eta}^2 + \lambda_4\abs{\Phi^\dagger\eta}^2 + \dfrac{\lambda_5}{2}\qty[(\Phi^\dagger\eta)^2 + \mathrm{h.c.}]+\tilde{\lambda}\abs{\Phi}^2\abs{S}^2+\lambda\abs{\eta}^2\abs{S}^2,
    \end{aligned}
    \label{eq:scalar_potential}
\end{equation}
where $\lambda_5$ can be taken as real by rephasing of the fields. 
We assume $\mu_1^2 < 0,\ \mu_S^2 < 0,\ \mu_2^2>0$, so that $\Phi$ and $S$ receive non-zero Vacuum Expectation Values (VEVs) by Spontaneous Symmetry Breaking (SSB) of the $\mathrm{SU(2)}_L\times\mathrm{U(1)}_Y$ and $\mathrm{U(1)}_{B-L}$ gauge symmetries. 
The VEV of $\eta$ remains zero.

The $\mathrm{U(1)}_{B-L}$ gauge symmetry is assumed to be spontaneously broken above the electroweak scale.
The scalar singlet $S$ is parameterized as 
\begin{align}
    S = \dfrac{1}{\sqrt{2}}\qty(v_S+\phi_S+iz_S),
\end{align}
where $v_S$ is the VEV of the $\mathrm{U}(1)_{B-L}$ symmetry breaking, $\phi_S$ is a neutral scalar boson, and $z_S$ is the Nambu--Goldstone (NG) boson.
The $Z^\prime$ boson obtains its mass as $m_{Z^\prime} = 2g_{B-L}v_S$ by its longitudinal mode absorbing $z_S$, where $g_{B-L}$ is the gauge coupling for the $\mathrm{U(1)}_{B-L}$ symmetry.
In the $\mathrm{SU(2)}_L\times\mathrm{U(1)}_Y\times\mathrm{U(1)}_{B-L}$ gauge sector, we suppose that the kinetic mixing between Z and $Z^\prime$ is negligible. 
RH neutrinos $N_\alpha$ also receive their masses as 
\begin{equation}
    m_{N_\alpha} = \dfrac{y^R_\alpha v_S}{\sqrt{2}}.
\end{equation}

After the Electro-Weak Symmetry Breaking (EWSB), the $\mathrm{SU}(2)_L$ scalar doublet $\Phi$ is parameterized as 
\begin{align}
    \Phi =
    \begin{pmatrix}
        G^+ \\
        \dfrac{1}{\sqrt{2}}\qty(v+\phi+iz)
    \end{pmatrix},
\end{align}
where $v$ ($\simeq 246$~GeV) is the VEV of the electroweak symmetry breaking, $G^+$ and $z$ are NG bosons absorbed by longitudinal modes of the electroweak gauge bosons, i.e., W and Z bosons, respectively. 
The stationary conditions give the following relations
\begin{align}
    \mu_1^2+\dfrac{\lambda_1}{2}v^2+\dfrac{\tilde{\lambda}v_S^2}{2} = 0, ~~~~\mu_S^2+\lambda_Sv_S^2+\dfrac{\tilde{\lambda}v^2}{2} = 0.
\end{align}
The mixing between $\phi$ and $\phi_S$ leads to the following mass terms
\begin{align}
\dfrac{1}{2}\begin{pmatrix}
        \phi & \phi_S
    \end{pmatrix}
    \mathcal{M}^2
    \begin{pmatrix}
        \phi\\
        \phi_S
    \end{pmatrix} =
    \dfrac{1}{2}\begin{pmatrix}
        \phi & \phi_S
    \end{pmatrix}
    \begin{pmatrix}
        \lambda_1v^2 & \widetilde{\lambda}vv_S\\
        \widetilde{\lambda}vv_S & 2\lambda_Sv_S^2
    \end{pmatrix}
    \begin{pmatrix}
        \phi\\
        \phi_S
    \end{pmatrix}.
\end{align}
The squared mass matrix $\mathcal{M}^2$ can be diagonalized by an orthogonal matrix with the mixing angle~$\alpha$
\begin{equation}
    \begin{pmatrix}
        h_1\\
        h_2
    \end{pmatrix}=
    \begin{pmatrix}
        \cos\alpha & -\sin\alpha\\
        \sin\alpha & \cos\alpha
    \end{pmatrix}
    \begin{pmatrix}
        \phi\\
        \phi_S
    \end{pmatrix},
\end{equation}
where $h_1$ and $h_2$ are mass eigenstates of Higgs bosons
\begin{align}
    &m_{h_1}^2 = \lambda_1v^2\cos^2\alpha+2\lambda_Sv_S^2\sin^2\alpha-\widetilde{\lambda}vv_S\sin2\alpha,\\
    &m_{h_2}^2 = \lambda_1v^2\sin^2\alpha+2\lambda_Sv_S^2\cos^2\alpha+\widetilde{\lambda}vv_S\sin2\alpha,
\end{align}
with the constraint
\begin{equation}
    \widetilde{\lambda}vv_S\cos2\alpha+\qty(\dfrac{\lambda_1}{2}v^2-\lambda_Sv_S^2)\sin2\alpha = 0.
\end{equation}
In this paper, we fix the mass eigenstate $h_1$ to be the SM-like Higgs boson with $m_{h_1} = 125$ GeV, and $h_2$ is the additional Higgs boson.

The $\mathbb{Z}_2$-odd scalar $\mathrm{SU(2)}_L$ doublet field $\eta$ can be parameterized as 
\begin{equation}
    \eta = \begin{pmatrix}
        H^+\\
        \dfrac{1}{\sqrt{2}}(H+iA)
    \end{pmatrix}.
\end{equation}
The mass spectrum of $\mathbb{Z}_2$-odd scalar particles is 
\begin{align}
    &m^2_{H^\pm} = \mu_2^2 + \dfrac{\lambda}{2}v_S^2 + \dfrac{\lambda_3}{2}v^2,\\
    &m^2_{H} = \mu_2^2 + \dfrac{\lambda}{2}v_S^2 + \dfrac{\lambda_3+\lambda_4+\lambda_5}{2}v^2,\\
    &m^2_{A} = \mu_2^2 + \dfrac{\lambda}{2}v_S^2 + \dfrac{\lambda_3+\lambda_4-\lambda_5}{2}v^2.
\end{align}

There are 11 parameters $\mu_1^2, \mu_2^2, \mu_S^2, \lambda_1, \lambda_2, \lambda_3, \lambda_4, \lambda_5, \lambda, \lambda_S, \tilde{\lambda}$ in the scalar potential Eq.(\ref{eq:scalar_potential}).
They can be replaced by 9 new physics parameters $\mu_2^2, m_{h_2},m_{H},m_{A},m_{H^\pm},\alpha,v_S,\lambda_2$ and $\lambda$, in addition to the two SM parameters $v$ and $m_{h_1}$.

\subsection{Neutrino masses}
\begin{figure}[t]
    \centering
    \includegraphics[scale=1.0]{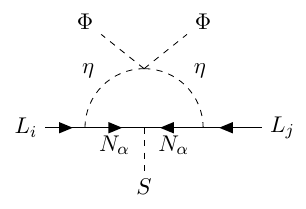}
    \caption{The one-loop diagram which generates small neutrino masses.}
    \label{fig: nu_mass}
\end{figure}

In this model, tiny neutrino masses are generated via the one-loop induced dimension-six operator $S\overline{(L\Phi)^c}L\Phi/\Lambda^2$, where $\Lambda$ is an energy scale parameter, as shown in  FIG.~\ref{fig: nu_mass}.

Following the framework of Ref.~\cite{Tao:1996vb, Ma:2006km}, after the EWSB, the mass matrix of light neutrinos at one-loop level is 
\begin{align}
    m_\nu^{ij} = \sum_\alpha g_{i\alpha}g_{j\alpha} \Lambda_{\alpha} \equiv (g\Lambda g^T)^{ij},
\end{align}
where the diagonal matrix $\Lambda$ is defined by
\begin{align}
    \Lambda_\alpha = \dfrac{m_{N_\alpha}}{32\pi^2}\qty[\dfrac{m_H^2}{m_{N_\alpha}^2-m_H^2}\ln\qty(\dfrac{m_{N_\alpha}^2}{m_H^2})-\dfrac{m_A^2}{m_{N_\alpha}^2-m_A^2}\ln\qty(\dfrac{m_{N_\alpha}^2}{m_A^2})].
\end{align}

In order to evaluate the Yukawa coupling of neutrinos, we adopt the Casas-Ibarra (CI) parametrization~\cite{Casas:2001sr}. 
The matrix of the Yukawa coupling can be parameterized as 
\begin{equation}
    g_{i\alpha} = \qty(U_\mathrm{PMNS}\sqrt{\mathcal{M}_\nu}R\sqrt{\Lambda^{-1}})_{i\alpha},
\end{equation}
where $U_{\mathrm{PMNS}}$ is the Pontecorvo-Maki-Nakagawa-Sakata matrix~\cite{Pontecorvo:1957qd, Maki:1962mu}, and the diagonalized mass matrix of active neutrinos is defined as $\mathcal{M}_\nu \equiv \mathrm{diag}(m_{\nu_1}, m_{\nu_2}, m_{\nu_3})$. 
The matrix $R$ is an arbitrary complex orthogonal matrix. We take $R = I$ in the following analysis.

\section{Constraints on the model}\label{sec:constraint}
In this section, we discuss possible theoretical and experimental constraints in this model. 
We consider perturbativity and vacuum stability as theoretical constraints. 
We consider experimental constraints from neutrino oscillations, collider experiments, flavor experiments, dark matter observations as well as the electroweak precision tests.

\subsection{Theoretical constraints}
In this subsection, we consider theoretical bounds in the model: perturbativity and vacuum stability. 
Perturbativity requires that all the quartic vertices of scalar fields should satisfy the perturbativity criterion $\lambda_1, \lambda_2, \lambda_3, \lambda_4, \lambda_5, \lambda_S,\widetilde{\lambda},\lambda < 4\pi$, and Yukawa couplings should satisfy $(y^R_1)^2, (y^R_2)^2, (y^R_3)^2 < 4\pi$.

Next, vacuum stability at tree level requires that the scalar potential is  bounded from below in all the directions of the field space~\cite{Deshpande:1977rw,Klimenko:1984qx,Sher:1988mj,Nie:1998yn,Kanemura:1999xf,Kanemura:2000bq,Ferreira:2004yd,Barger:2008jx,Ginzburg:2010wa,Fuyuto:2014yia}.
Therefore, the vacuum stability requires
\begin{align}
    &\lambda_1\geq 0,\ \lambda_2\geq 0,\ \lambda_3\geq-\sqrt{\lambda_1\lambda_2},\ \lambda_3+\lambda_4\pm\lambda_5\geq-\sqrt{\lambda_1\lambda_2},\notag\\
    &\lambda_S\geq0,\ \lambda\geq-\sqrt{2}\sqrt{\lambda_2\lambda_S},\ \tilde{\lambda}\geq-\sqrt{2}\sqrt{\lambda_1\lambda_S}.
\end{align}

\subsection{Constraints from experiments}
First, we discuss constraints from neutrino oscillation.  
We consider normal mass ordering of neutrinos, and adopt the latest data from the PDG~\cite{ParticleDataGroup:2022pth}, so our analysis is compatible with neutrino oscillation measurements:
\begin{align}
    &\Delta m_{21}^2 = (7.53\pm0.18)\times10^{-5}\ \mathrm{eV}^2, ~\Delta m_{23}^2 = (2.455\pm0.028)\times10^{-3}\ \mathrm{eV}^2, ~ \delta_{CP} = (1.19\pm0.22)\pi, \notag\\
    &\sin^2(\theta_{12}) = 0.307\pm0.013, ~ \sin^2(\theta_{13}) = 0.0219\pm0.0007, ~\sin^2(\theta_{23}) = 0.558\pm0.015.
\end{align}

Next, we consider constraints from lepton flavor violation (LFV). 
In this model, LFV decay processes can be achieved via one-loop diagrams mediated by $H^\pm$ and $N_\alpha$. 
We consider constraints from $\ell_i\to\ell_j\gamma$ and $\ell_i\to\ell_j\ell_k\bar{\ell_k}$, as shown in Table~\ref{tab:clfv_bound}.
\begin{table}[t]
    \centering
    \begin{tabular}{|c|c|}
        \hline
        LFV processes & Current bounds on branching ratios \\
        \hline
        $\mu^+\to e^+\gamma$ & $3.1\times10^{-13}$~\cite{MEGII:2023ltw}\\
        \hline
        $\mu^+\to e^-e^+e^-$ & $1.0\times10^{-12}$~\cite{SINDRUM:1987nra}\\
        \hline
        $\tau^+\to e^+\gamma$ & $3.3\times10^{-8}$~\cite{BaBar:2009hkt}\\
        \hline
        $\tau^+\to \mu^+\gamma$ & $4.2\times10^{-8}$~\cite{Belle:2021ysv}\\
        \hline
        $\tau^-\to e^-e^+e^-$ & $2.7\times10^{-8}$~\cite{Hayasaka:2010np}\\
        \hline
        $\tau^-\to e^+\mu^-\mu^-$ & $1.7\times10^{-8}$~\cite{Hayasaka:2010np}\\
        \hline
        $\tau^-\to e^-\mu^+\mu^-$ & $2.7\times10^{-8}$~\cite{Hayasaka:2010np}\\
        \hline
        $\tau^-\to \mu^+e^-e^-$ & $1.5\times10^{-8}$~\cite{Hayasaka:2010np}\\
        \hline
        $\tau^-\to \mu^-e^+e^-$ & $1.8\times10^{-8}$~\cite{Hayasaka:2010np}\\
        \hline
        $\tau^-\to \mu^-\mu^+\mu^-$ & $2.1\times10^{-8}$~\cite{Hayasaka:2010np}\\
        \hline
        $\mu^-\mathrm{Ti}\to e^-\mathrm{Ti}$ & $4.3\times10^{-12}$~\cite{SINDRUMII:1993gxf}\\
        \hline
        $\mu^-\mathrm{Au}\to e^-\mathrm{Au}$ & $7.0\times10^{-13}$~\cite{SINDRUMII:2006dvw}\\
        \hline
    \end{tabular}
    \caption{Current experimental bounds for LFV processes.}
    \label{tab:clfv_bound}
\end{table}

Third, this model basically provides two candidates of DM, which are the lightest RH neutrino or the lightest neutral $\mathbb{Z}_2$ odd scalar particle.  
The relic density of DM candidates should satisfy observation results from the Planck satellite $\Omega h^2 = 0.120\pm 0.001$~\cite{Planck:2018vyg}. 
Direct detection (DD) experiments can also give constraints to this model. 
We apply limits from the LZ experiment~\cite{LZ:2022lsv} on the WIMP-nucleon spin independent elastic scattering cross section. 
Detailed analysis is shown in the next section.

Fourth, collider experiments give rich constraints on this model, including masses of new particles (e.g. $Z^\prime$, $m_{h_2}$, inert scalar particles) and BSM parameters (e.g. $\alpha$ and $v_S$). 
The LEP II experiment gives the lower limit of the ratio between the mass of $Z^\prime$ gauge boson and its coupling constant $m_{Z^\prime}/g_{B-L} > 7$~TeV~\cite{Electroweak:2003ram,Carena:2004xs,Cacciapaglia:2006pk}. 
Chargino searches at the LEPII experiment can provide a limit of $m_{H^\pm}>80$~GeV~\cite{ALEPH:2013htx}. 
The LEPII experiment also exclude the intersection of the following mass region for inert scalar particles~\cite{Lundstrom:2008ai}:
\begin{equation}
    m_H < 80\ \mathrm{GeV},\ m_A < 100\ \mathrm{GeV},\ m_A-m_H > 8\ \mathrm{GeV}.
\end{equation}
With precise measurements of decay widths of W and Z bosons~\cite{ParticleDataGroup:2022pth}, kinematically allowed regions for decay processes $W/Z\to\eta\eta$ are excluded.
The null result in searching for the $Z^\prime$ boson at ATLAS~\cite{ATLAS:2019erb,ATLAS:2019fgd} and CMS~\cite{CMS:2021ctt,CMS:2022eud} gives strong constraints on the $Z^\prime$ boson, $m_{Z^\prime} < 5.15\ \mathrm{TeV}$.  
Precise measurements for the Higgs boson couplings give constraints on the mixing between $\phi$ and $\phi_S$.
The scaling factor for the $h_1ZZ$ coupling is given by 
\begin{equation}
    \kappa_Z^2\equiv\dfrac{\Gamma(h_1\to ZZ)}{\Gamma_\mathrm{SM}(h_1\to ZZ)} = \cos^2\alpha,
\end{equation} 
where ATLAS~\cite{ATLAS:2022vkf} and CMS~\cite{CMS:2022dwd} provide 
\begin{equation}
    \kappa_Z = 1.04 \pm 0.07,
\end{equation}
from which we obtain $0.90 < \cos\alpha \le 1.0$ as a criterion of 95\% CL~\footnote{The allowed parameter space for $m_{h_2}$ and $\alpha$ is shown in Ref.~\cite{Robens:2015gla,Dupuis:2016fda} from the direct search results for $h_2$.}. 

Finally, we will consider constraints from the electroweak precision test. 
Electroweak precision measurements can provide strong constraints on BSM physics. 
The oblique parameter $T$ gives the strongest constraint on a multi-Higgs scenario among $S,T,U$ parameters~\cite{Peskin:1990zt,Peskin:1991sw}.
In this model, inert doublet field and scalar singlet field contribute to deviations of $T$~\cite{Toussaint:1978zm,Barger:2007im,Kanemura:2011sj}:
\begin{align}
    \Delta T &= \dfrac{3\sin^2\alpha}{16\pi s_W^2}\qty[f_T\qty(\frac{m_{h_2}^2}{m_W^2})-f_T\qty(\frac{m_{h_1}^2}{m_W^2})-\dfrac{1}{c_W^2}\qty(f_T\qty(\frac{m_{h_2}^2}{m_Z^2})-f_T\qty(\frac{m_{h_1}^2}{m_Z^2}))] \notag \\
    &+ \dfrac{1}{16\pi m_W^2 s_W^2}\qty[F(m_{H^\pm},m_H)+F(m_{H^\pm},m_A)-F(m_H,m_A)],
\end{align}
where $c_W^2\equiv m_W^2/m_Z^2$ and $s_W^2=1-c_W^2$.
The auxiliary functions are defined as 
\begin{align}
    F(x,y) = \dfrac{x^2+y^2}{2}-\dfrac{x^2y^2}{x^2-y^2}\ln\dfrac{x^2}{y^2},~~~~ f_T(x)=\dfrac{x\log x}{x-1}.
\end{align}

\section{Lepton flavor violation}\label{sec:CLFV}
In this section, we consider constraints from LFV. 
LFV processes can be enhanced through $N_\alpha$--$H^\pm$ loop diagrams in this model. 
The branching ratio for $\ell_i\to\ell_j\gamma$ is calculated as
\begin{equation}
    \mathrm{Br}(\ell_i\to\ell_j\gamma) = \dfrac{48\pi^3\alpha_{\mathrm{em}}\abs{A_D}^2}{G_F^2}\mathrm{Br}(\ell_i\to\ell_j\overline{\nu}_j\nu_i),
\end{equation}
with
\begin{equation}
    A_D = \sum_{\alpha}\dfrac{ig_{i\alpha}^*g_{j\alpha}}{32\pi^2m_{H^\pm}^2}F(\xi_\alpha),
\end{equation}
where $\xi_\alpha$ is defined as $\xi_\alpha \equiv m_{N_\alpha}^2/ m_{H^\pm}^2$, $\alpha_{\mathrm{em}}$ is the fine structure constant, $G_F$ is the Fermi constant, and the function $F(x)$ is defined as 
\begin{equation}
    F(x) = \dfrac{1-6x+3x^2+2x^3-6x^2\ln x}{6(x-1)^4}.
\end{equation} 
$\ell_i\to\ell_j\ell_k\bar{\ell_k}$ processes include photon penguin diagrams and box diagrams. The branching ratios of photon penguin diagrams can have a simple relation with the radiative decay processes~\cite{Arganda:2005ji,Ilakovac:2012sh,Toma:2013zsa}:
\begin{align}
    &\mathrm{Br}(\ell_i\to\ell_j\ell_j\bar{\ell_j})\simeq\dfrac{\alpha_\mathrm{em}}{3\pi}\qty[2\log(\dfrac{m_{\ell_i}}{m_{\ell_j}})-\dfrac{11}{4}]\mathrm{Br}(\ell_i\to\ell_j\gamma),\\
    &\mathrm{Br}(\ell_i\to\ell_j\ell_k\bar{\ell_k})\simeq\dfrac{\alpha_\mathrm{em}}{3\pi}\qty[2\log(\dfrac{m_{\ell_i}}{m_{\ell_j}})-3]\mathrm{Br}(\ell_i\to\ell_j\gamma)~~(j\neq k).
\end{align}
The box diagrams can be dominant when $\xi\gg1$ or $\xi\ll1$ if masses of RH neutrinos are degenerate.
$\mu\to e$ conversion in muonic atoms is also a signature of LFV. However, $\mu\to e\gamma$ is more stringent than this process~\cite{Toma:2013zsa}.
Detailed analysis of LFV can be seen in Ref.~\cite{Arganda:2005ji,Ilakovac:2012sh,Toma:2013zsa,Ardu:2022sbt}.

\begin{figure}[t]
    \centering
    \includegraphics[scale = 0.7]{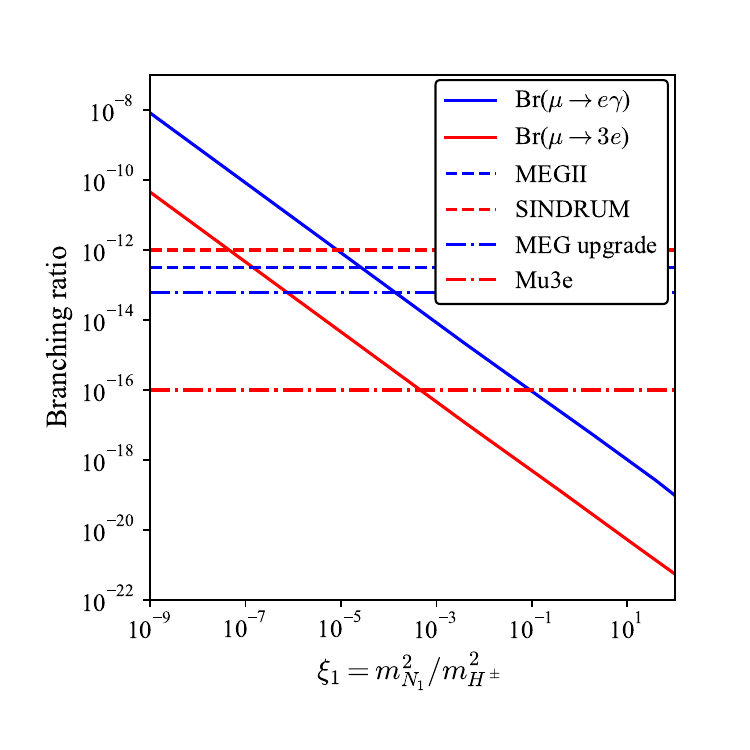}
    \caption{$\mathrm{Br}(\mu\to e\gamma)$ (blue) and $\mathrm{Br}(\mu\to 3e)$ (red) as a function of $\xi = m_{N_1}^2/m_{H^\pm}^2$. Horizontal dashed lines show the current upper bounds from MEGII experiment (blue)~\cite{MEGII:2023ltw} and SINDRUM experiment (red)~\cite{SINDRUM:1987nra}. The dot-dashed lines show the future bounds from MEG upgrade (blue)~\cite{Baldini:2013ke} and Mu3e (red)~\cite{Blondel:2013ia}.}
    \label{fig:lfv1}
\end{figure}
Considering the constraints from neutrino oscillations and LFV measurements, we choose the following parameters for later use:
\begin{align}
    &m_{N_2} = m_{N_1} + 2500 \ \mathrm{GeV},\ m_{N_3} = m_{N_1} + 3000 \ \mathrm{GeV},\ m_H = 1000 \ \mathrm{GeV},\notag \\
    &m_{H^\pm} = m_A,\ \delta\equiv m_{H^\pm} - m_H = 10^{-5}\ \mathrm{GeV}.
    \label{eq:base_bench}
\end{align}

LFV constraints on the mass of $N_1$ is shown in FIG.~\ref{fig:lfv1}. 
The blue (red) solid line shows $\mathrm{Br}(\mu \to e \gamma)$ ($\mathrm{Br}(\mu \to 3 e)$).
The horizontal dashed blue and red lines are current upper bounds from MEGII~\cite{Baldini:2013ke} and SINDRUM~\cite{Blondel:2013ia}, respectively.
Future expected bounds from MEG upgrade~\cite{Baldini:2013ke} and Mu3e~\cite{Blondel:2013ia} are shown as dot-dashed lines.
As $m_{N_1}$ is increasing, branching ratios of LFV processes are decreasing, which provides the lower bound of $m_{N_1}$ around $m_{N_1} \simeq 5.1$ GeV.

\section{Dark matter}\label{sec:DM}
In this section, we consider the freeze-out mechanism for DM candidates. 
The relic abundance of the DM particle can be calculated by solving the Boltzmann equation~\cite{Gondolo:1990dk}:
\begin{equation}
    \dfrac{\mathrm{d}Y}{\mathrm{d}x} = -\sqrt{\dfrac{g_*\pi}{45}}\dfrac{m_\mathrm{DM}m_{Pl}}{x^2}\expval{\sigma v}(Y^2-Y_{EQ}^2),
\end{equation}
where $x\equiv m_\mathrm{DM}/T$ is an independent variable, $Y\equiv n_\mathrm{DM}/s$ is the comoving density of DM. 
The thermal averaged cross section $\expval{\sigma v}$ contains cross sections of all annihilation processes before the DM freezes out. 
Integrating the Boltzmann equation from $x = 0$ to $x_F = m_\mathrm{DM}/T_F$, the relic abundance is given by
\begin{equation}
    \Omega_\mathrm{DM}h^2 \simeq 2.75\times10^{-8}m_\mathrm{DM}Y(T_F)\mathrm{GeV}^{-1},
\end{equation}
where $T_F$ is the freeze-out temperature of DM.

In this paper, we discuss two scenarios for two DM candidates.
We first discuss fermionic DM scenario where $N_1$ is the lightest particle in the $\mathbb{Z}_2$ odd sector, and show allowed parameter spaces under current various experiments.
We then briefly discuss scalar DM scenario in the model, whose discussions are similar to several previous studies~\cite{Lundstrom:2008ai,Belyaev:2016lok,LopezHonorez:2006gr,DuttaBanik:2014iad,Khojali:2022squ}.
In the following analysis, we use the numerical analysis package micrOMEGAs 6.0~\cite{Alguero:2023zol} to calculate the relic abundance of DM. 

\subsection{$N_1$ dark matter}
\begin{figure}[t]
    \centering
    \includegraphics[scale=1.0]{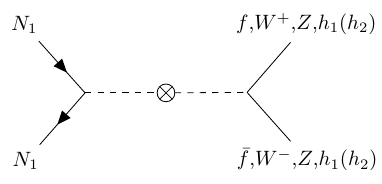}
    \caption{Annihilation processes $N_1N_1\to h_1(h_2)\to f\bar{f},ZZ,W^+W^-,h_1(h_2)h_1(h_2)$ through the mixing between $\phi$ and $\phi_S$.
    }
    \label{fig: mixing}
\end{figure}
In this scenario, we consider $N_1$ as the lightest $\mathbb{Z}_2$ odd particle. 

First, we consider the relic abundance of $N_1$ dark matter. 
In the Tao-Ma model~\cite{Tao:1996vb,Ma:2006km}, there is no sufficient annihilation rates for $N_1$ DM due to strong constraints from LFV experiments~\cite{Kubo:2006yx}. 
However, it is possible for $N_1$ to satisfy the DM abundance with the mixing between the Higgs field $\Phi$ and the extra scalar singlet $S$, because $N_1$ can annihilate via $N_1N_1\to h_1(h_2)\to f\bar{f},ZZ,W^+W^-,h_1(h_2)h_1(h_2)$ processes~\cite{Okada:2010wd}, as shown in FIG.~\ref{fig: mixing}.

\begin{figure}[t]
    \centering
    \includegraphics[scale = 0.9]{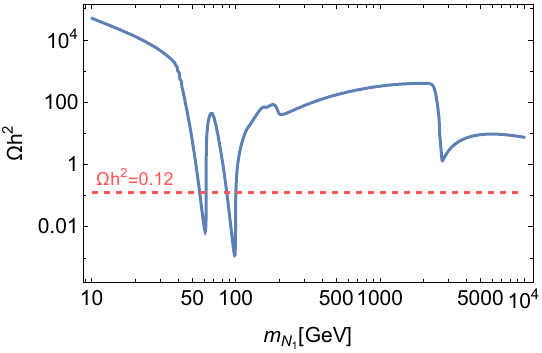}
    \caption{Relic abundance of $N_1$ as a function of $m_{N_1}$. 
    The red dashed line is the current bound of DM relic abundance from the Planck experiment~\cite{Planck:2018vyg}.}
    \label{fig:relic_abundance}
\end{figure}

In FIG.\ref{fig:relic_abundance}, the relic abundance of $N_1$ is shown as a function of $m_{N_1}$. 
We use the following parameters to calculate the relic abundance:
\begin{align}
    &\cos\alpha = 0.97,\ v_S = 30\ \mathrm{TeV},\ m_{h_2} = 200\ \mathrm{GeV},\ m_{Z^\prime}=5.2\ \mathrm{TeV},\notag \\
    &m_{N_2} = m_{N_1} + 2500\ \mathrm{GeV},\ m_{N_3} = m_{N_1} + 3000\ \mathrm{GeV},\ \lambda = 10^{-6},\ \lambda_3 = 0.1,\notag \\
    &m_H = m_{N_1} +10^3\ \mathrm{GeV},\ m_{H^\pm} = m_A = m_H + 10^{-5}\ \mathrm{GeV}.
    \label{eq:para_set}
\end{align}
The relic abundance of $N_1$ is reduced significantly around $m_{N_1}\simeq m_{h_1}/2$ and $m_{N_1}\simeq m_{h_2}/2$, respectively. 
Annihilations of RH neutrinos to the SM particles are enhanced resonantly via $s$-channel exchange of $h_1$ and $h_2$. 
We also notice that it is possible for RH neutrinos to annihilate via $Z^\prime$ boson exchange. 
However, this process is not dominant since the cross section of $Z^\prime$ boson exchange $\expval{\sigma v}$ is proportional to $1/v_S^4$, which is much smaller than Higgs boson exchange processes in the low mass region $m_{N_1}\ll m_{Z^\prime}/2$. 
In the mass region near the $Z^\prime$ pole $m_{N_1}\simeq m_{Z^\prime}/2$, even if annihilations can be enhanced due to $Z^\prime$ resonance, it cannot realize the required relic abundance.

Second, we consider the direct detection (DD) of $N_1$ dark matter. 
The lightest RH neutrino $N_1$ can have elastic scattering with nucleons by Higgs exchange processes. 
The spin independent (SI) cross section for the proton target is 
\begin{equation}
    \sigma_{\mathrm{SI}}^{p} = \dfrac{4\mu^2}{\pi}f_p^2,
\end{equation}
where $\mu\equiv\dfrac{m_{N_1}m_p}{m_{N_1}+m_p}$ is the DM-nucleus reduced mass in the center of mass frame. The hadronic matrix element $f_p$~\cite{DelNobile:2021wmp} is given by 
\begin{equation}
    f_p = \sum_{q=u,d,s}c_q\dfrac{m_p}{m_q}f_{Tq}^p + \dfrac{2}{27}f_{TG}^p\sum_{q=c,b,t}c_q\dfrac{m_p}{m_q}, \label{eq:fp}
\end{equation}
where $f_{Tq}^p$ and $f_{TG}^p$ express mass contributions to the nucleon from quarks and gluons, and $m_q$ is the mass of a quark with a Yukawa coupling $y_q$. The effective vertex $c_q$ is defined as 
\begin{equation}
    c_q = y_R^\alpha y_q\bigg[\qty(\dfrac{\sin\alpha}{\sqrt{2}})\dfrac{1}{m_{h_1}^2}\qty(\dfrac{\cos\alpha}{\sqrt{2}})-\qty(\dfrac{\sin\alpha}{\sqrt{2}})\dfrac{1}{m_{h_2}^2}\qty(\dfrac{\cos\alpha}{\sqrt{2}})\bigg].
\end{equation}
The spin independent cross section $\sigma_{\mathrm{SI}}^{p}$ has a simple relation with BSM parameters in this model:
\begin{equation}
    \sigma_{\mathrm{SI}}^{p}\propto\qty(\dfrac{m_{N_1}\sin(2\alpha)}{v_S})^2\times\qty(\dfrac{1}{m_{h_1}^2}-\dfrac{1}{m_{h_2}^2})^2.
\end{equation}

\begin{figure}[t]
    \centering
    \includegraphics[scale = 0.8]{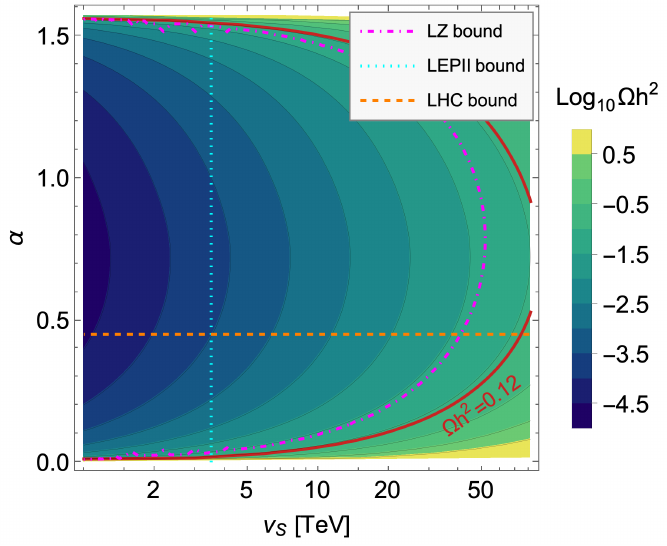}
    \caption{Parameter space of $m_{N_1}=100.1\ \mathrm{GeV}$ with constraints from LEPII~\cite{Electroweak:2003ram,Carena:2004xs}, ATLAS~\cite{ATLAS:2022vkf}, CMS~\cite{CMS:2022dwd} and LZ~\cite{LZ:2022lsv} experiments. The red line is the bound from the Planck experiment~\cite{Planck:2018vyg}.}
    \label{fig:mN1_100_2}
\end{figure}

In FIG.~\ref{fig:mN1_100_2}, the relic abundance of $N_1$ DM is shown as colored contour in the $v_S$-$\alpha$ plane. We choose the following parameter set:
\begin{align}
        &m_{N_1} = 100.1\ \mathrm{GeV},\ m_{h_2} = 200\ \mathrm{GeV},\ m_{N_2} = 2500 \ \mathrm{GeV},\ m_{N_3} = 3000 \ \mathrm{GeV},\notag \\
        &\lambda=10^{-6},\ \lambda_3=0.1,\ m_H = 1000\ \mathrm{GeV},\ m_{H^\pm} = m_A,\ \delta\equiv m_{H^\pm} - m_H = 10^{-5}\ \mathrm{GeV}.
        \label{eq:para_sets}
\end{align}
The allowed parameter space is the region encircled by the LZ 2022~\cite{LZ:2022lsv} (magenta dot-dashed), LHC~\cite{ATLAS:2022vkf,CMS:2022dwd} (orange dashed) and Planck experiment~\cite{Planck:2018vyg} (red solid), which is on the right side of the LEPII constraint~\cite{Electroweak:2003ram,Carena:2004xs} (cyan dotted). 
We also investigate the parameter space of the SM-like Higgs resonance near the region $m_{N_1}\simeq m_{h_1}/2$. 
We find that there are still parameter spaces for $N_1$ DM with the SM-like Higgs resonance if $m_{h_2}$ is near $m_{h_1}$.

\begin{figure}[t]
    \centering
    \includegraphics[scale = 1.0]{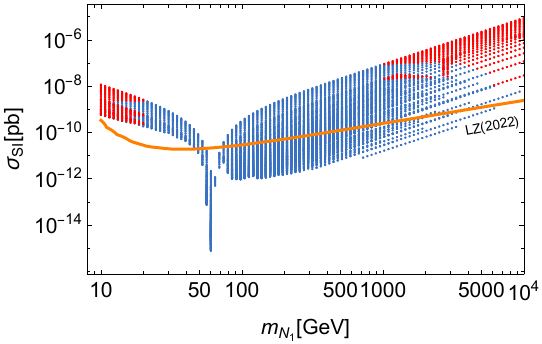}
    \caption{Possible SI cross section of $N_1$ as a function of $m_{N_1}$ in our parameter space. Red points are excluded by theoretical constraints. The orange line is the upper bound from the LZ experiment~\cite{LZ:2022lsv}.}
    \label{fig:para_scan}
\end{figure}

Third, we explore the allowed mass region of $N_1$ with $h_2$ resonance scenario.
We choose the following parameter space 
\begin{equation}
    \cos \alpha \in [0.9,1],\ v_S \in [3.5, 80]\ \mathrm{TeV},\ m_{h_2}\in[10 , 10^4]\ \mathrm{GeV}
\end{equation}
with 
\begin{align}
    &m_{N_2} = m_{N_1} + 2500\ \mathrm{GeV},\ m_{N_3} = m_{N_1} + 3000\ \mathrm{GeV},\ m_{h_2} = 2m_{N_1} + 0.4\ \mathrm{GeV},\notag \\
    &\lambda = 10^{-6},\ \lambda_3 = 0.1,\ m_H = m_{N_1} +10^3\ \mathrm{GeV},\ m_{H^\pm} = m_A = m_H + 10^{-5}\ \mathrm{GeV}.
\end{align}
We scan $\cos \alpha$, $v_S$ and $m_{h_2}$ with $20 \times 20 \times 100$ points, respectively.
The allowed parameter space of $N_1$ is shown in FIG.~\ref{fig:para_scan}, in which all points satisfy the constraint $\Omega_{\mathrm{DM}}h^2 < 0.12$.
Blue points satisfy all theoretical and experimental constraints, while red points are excluded by theoretical constraints.
The low mass region $m_{N_1} \lesssim 45$~GeV is excluded by the LZ experiment since the SI cross section rapidly increases due to the large contribution of $h_2$ exchange processes. 
In the mass region near $m_{N_1}\simeq m_{h_1}/2$, the SI cross section greatly reduces due to the destructive interference of the two Higgs states. 
In high mass regions $m_{N_1} \gtrsim 80 ~\mathrm{GeV}$, the SI cross section increases as $m_{N_1}$ is increasing. However, there is still allowed mass region for $N_1$ DM. 
In the region $m_{N_1} > 5~\mathrm{TeV}$, the SI cross section exceeds the LZ bound again, therefore current direct detection experiments can give the lower and upper bounds on the mass of $N_1$.

\subsection{Scalar dark matter}
In this scenario, we consider $H$ is the lightest $\mathbb{Z}_2$ odd particle. 
The $H$ DM scenario has been studied extensively in the inert doublet model~\cite{Lundstrom:2008ai,LopezHonorez:2006gr,Aoki:2013lhm,Kanemura:2016sos,Belyaev:2016lok}.
In our model, there are two differences with the inert doublet model. 
Discussions of $H$ DM is similar to the inert doublet model with an additional scalar singlet~\cite{DuttaBanik:2014iad,Khojali:2022squ}.

First, the relic density can be satisfied by the additional Higgs resonance near the mass region $m_{H}\simeq m_{h_2} /2$. 
$H$ can annihilate via $HH\to h_1(h_2)\to f\bar{f},W^+W^-,ZZ,h_1(h_2)h_1(h_2)$ processes. 

Second, the SI cross section in direct detections can be changed due to the $h_2$ exchange processes. 
The SI cross section of $H$ is given by
\begin{equation}
    \sigma_{\mathrm{SI}}^{H} = \dfrac{\mu_H^2}{4\pi m_H^2}f_p^2,\ \mu_H \equiv \dfrac{m_Hm_p}{m_H+m_p},
\end{equation}
where $f_p$ has the same definition as Eq.~(\ref{eq:fp}), and $c_q$ is defined as~\footnote{The SI cross section can be modified depending on the size of $\lambda_2$ via loop corrections~\cite{Abe:2015rja}.}
\begin{align}
    c_q = \dfrac{y_q}{2\sqrt{2}}\qty[\dfrac{\lambda v_S\sin\alpha\cos\alpha-\lambda_Lv\cos^2\alpha}{m_{h_1}^2}-\dfrac{\lambda_Lv\sin^2\alpha+\lambda v_S\sin\alpha\cos\alpha}{m_{h_2}^2}].
\end{align}

We here take a benchmark point for the $H$ DM scenario:
\begin{align}
    &v_S = 5\ \mathrm{TeV},\ \mu_2^2 = 3460\ \mathrm{GeV}^2,\ m_{h_2} = 120 \ \mathrm{GeV},\ \cos\alpha=0.98,\ \lambda = 10^{-4},\ m_{H^\pm} = 100\ \mathrm{GeV},\notag\\ 
    &m_{H} = 60\ \mathrm{GeV},\ m_{A} = 100\ \mathrm{GeV},\ m_{N_1} = 3000\ \mathrm{GeV},\ m_{N_2} = 3500\ \mathrm{GeV},\ m_{N_3} = 4000\ \mathrm{GeV}.
\end{align}
The SI cross section and the LZ value of this point are given by  
\begin{equation}
    \sigma_\mathrm{SI}^H = 5.35\times10^{-12}\ \mathrm{pb},\ \sigma_\mathrm{LZ(2022)} = 2.01\times10^{-11}\ \mathrm{pb}.
\end{equation}
The relic abundance of this benchmark point is evaluated to be 
\begin{equation}
    \Omega_{\mathrm{DM}} h^2 \simeq 0.12,
\end{equation}
indicating that there is still allowed parameter space for the $H$ DM scenario in this model.

\section{Discussions}\label{sec:discussions}
In this section, we give some comments on the results shown in section~\ref{sec:CLFV} and~\ref{sec:DM}. 
In section~\ref{sec:CLFV}, we have shown that experiment results give the lower bound of $m_{N_1}$. 
At present, the most stringent constraint comes from the $\mu\to e\gamma$ process, which gives the lower bound on $m_{N_1} > 5.2$~GeV in our $N_1$ DM scenario. 
In the future, the MEG experiment~\cite{Baldini:2013ke} may  improve its sensitivity down to $\mathrm{Br}(\mu\to e\gamma)\simeq6\times10^{-14}$, and the Mu3e experiment~\cite{Blondel:2013ia} may upgrade the sensitivity to $\mathrm{Br}(\mu\to 3e)\simeq1\times10^{-16}$. 
Future LFV experiments would give the lower bound $m_{N_1} > 21$~GeV in our $N_1$ DM scenario, which is 4 times better than the current bound.

In section~\ref{sec:DM}, we have shown benchmark points and parameter space of DM candidates in this model. 
Knowing the allowed parameter spaces, we can discuss the testability of new particles at future collider experiments. 
We here very briefly mention the prospect for collider searches in each scenario.
In the $N_1$ DM scenario, $N_1$ could be tested through $H^\pm\to l^\pm N_1$ signals at LHC and HL-LHC if it is kinematically allowed, mainly via $pp\to H^+H^-\to N_1N_1l^+l^-$~\cite{Moretti:2001pp,Alves:2005kr,Aoki:2013lhm,Belyaev:2016lok,Kalinowski:2018ylg,Kanemura:2021dez}.
In the $H$ DM scenario, inert scalar particles can be produced through monojet processes (e.g., $q\bar{q}\to Z g\to HAj$ and $q\bar{q}\to h_1g\to HHg$) in hadron colliders. 
$A$ and $H^\pm$ can decay to $H$ via weak interactions. 
Cross sections of processes including the $Z H A$ vertex only depend on $m_H$ and $m_A$. 
Cross section of processes including the $h_1HH$ vertex not only depend on $m_H$, but also the dimensionless parameter $\lambda_3 + \lambda_4 + \lambda_5$. 
Detailed analysis of inert scalars in the LHC can be found in Ref.~\cite{Alves:2005kr,Belyaev:2016lok,Kanemura:2021dez}. 
Inert scalars can also be produced in lepton colliders through $e^+e^-\to Z \to AH(H^+H^-)$ processes~\cite{Aoki:2013lhm,Kalinowski:2018ylg}. 

The existence of the $Z^\prime$ boson is another difference from the Tao-Ma model. 
Though searches in the LHC give strong constraints on the mass of $Z^\prime$ boson, it can be lighter with smaller gauge couplings. 
Since hadron colliders (e.g., LHC and HL-LHC) have large backgrounds, it can only reach the gauge coupling $g_{B-L}$ of $\mathcal{O}(10^{-2})$~\cite{Basso:2009hf}. 
However, lepton colliders (e.g., ILC) can reach smaller gauge coupling of $\mathcal{O}(10^{-3})$~\cite{Basso:2009hf}. 
Decays of the $Z^\prime$ boson into SM particles are proportional to $(B-L)^2$. 
Due to its large mass, the branching ratios of $Z^\prime\to q\bar{q}, l^+l^-, \nu_L\bar{\nu}_L,N_1 N_1,h_1h_1,h_1h_2$ and $h_2h_2$ for the parameter set Eq.~(\ref{eq:para_set}) are given approximately by 0.20, 0.29, 0.15, 0.15, $6.9\times10^{-4}$, $4.4\times10^{-2}$, 0.17,  respectively~\cite{Carena:2004xs,Kanemura:2011vm}.

Direct detection experiments become more and more stringent for DM candidates. The LZ collaboration has released its preliminary result with 280 days of data in 2024~\cite{LZ2024}, which has improved a lot compared to the LZ 2022 result~\cite{LZ:2022lsv}. Future DM detection experiments may give stronger bounds on the DM mass in this model.

\section{Conclusions}\label{sec:conclusion}
The radiative seesaw model with gauged $\mathrm{U(1)}_{B-L}\times\mathbb{Z}_2$ extension is a well-motivated scenario which gives consistent predictions of active neutrino masses and the abundance of dark matter. 
Majorana masses of right-handed neutrinos, the lightest of which can be identified as dark matter, are given by the spontaneous breaking of the $\mathrm{U(1)}_{B-L}$ gauge symmetry. 
We have revisited this model with the latest constraints from dark matter searches, neutrino oscillations, flavor experiments and collider experiments. 
We have explored the feasible parameter space of this model, and have found that there is still allowed region for this model under the latest experimental constraints. 
We have presented new viable benchmark scenarios for this model, i.e., the right-handed neutrino dark matter scenario and the scalar dark matter scenario. 
We also have mentioned the testability of these benchmark scenarios at future experiments.

\section*{Acknowledgements}
The work of S.~K. was supported by the JSPS Grants-in-Aid for Scientific Research No.~20H00160, No.~23K17691 and No.~24KF0060.
The work of Y.~M. was supported by the JSPS Grant-in-Aid for JSPS Fellows No.~23KJ1460.
The work of G.~Y. was supported by the PhD support program for Pioneering Quantum Beam Applications (PQBA) and the Nishimura International Scholarship foundation.

\bibliographystyle{unsrt}
\bibliography{ref}
\end{document}